\documentclass[a4paper,11pt]{article}

\usepackage{pos}

\title{Results from IceCube Follow-up of Nearby
Supernova SN2023ixf}

\ShortTitle{IceCube Follow-up of SN2023ixf}

\author{The IceCube Collaboration \\{\normalsize \normalfont(a complete list of authors can be found at the end of the proceedings)}\\}

\emailAdd{aemand@wisc.edu}
\emailAdd{justin.vandenbroucke@wisc.edu}
\emailAdd{jessie.thwaites@icecube.wisc.edu}
\emailAdd{sahori@icecube.wisc.edu}
\emailAdd{aswathi.balagopalv@icecube.wisc.edu}

\abstract{

Core-collapse supernovae are of particular interest in multi-messenger astronomy due to their potential to accelerate cosmic rays and produce high-energy neutrinos. One such supernova is the recent SN2023ixf located in M101 (the Pinwheel Galaxy). It is the closest (6.4 Mpc) and brightest (B band magnitude 10.8) core-collapse supernova in nearly a decade. This supernova likely had a progenitor surrounded by dense circumstellar material which, during the supernova, may have produced neutrinos when ejecta collided with the material. I will present results of a follow-up of this supernova using data collected from the IceCube Neutrino Observatory located at the South Pole. We obtain results consistent with background expectations with time-integrated energy flux (E$^2$ dN/dE) upper limits of 0.35 GeV/cm$^2$ for a 32-day time window and 0.44 GeV/cm$^2$ for a 4-day time window, both at 90\% confidence level for an E$^{-2}$ power law. These correspond to values of 2.7$\times 10^{48}$ erg for the 32-day time window and 3.5$\times 10^{48}$ erg for the 4-day time window at the supernova.

\vspace{4mm}

{\bfseries Corresponding authors:}
Alicia Mand$^{1*}$, 
Justin Vandenbroucke$^{1}$, 
Jessie Thwaites$^{1}$,
Sam Hori $^{1}$, 
Aswathi Balagopal V$^{2}$ \\
{$^{1}$ \itshape Dept. of Physics and Wisconsin IceCube Particle Astrophysics Center, University of Wisconsin—Madison}\\
{$^{2}$ \itshape University of Delaware}\\
$^*$ Presenter
}

\ConferenceLogo{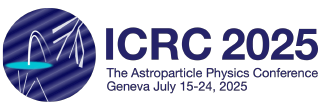}

\FullConference{39th International Cosmic Ray Conference (ICRC2025)\\
 15–24 July 2025\\
Geneva, Switzerland\\}

\begin{document}

\maketitle
\section{Introduction}\label{sec1}

Multi-messenger astronomy is a powerful tool for probing extreme astrophysical phenomena in new ways. By combining information from neutrinos and optical observations, we can examine high-energy transients with a new lens. One example of these transients are type-II core-collapse supernovae which produce optical light and are expected to produce high-energy neutrinos \cite{guetta_low-_2023}. 

Supernova 2023ixf (hereafter, SN 2023ixf) is a type-II core-collapse supernova discovered on May 19, 2023 by K. Itagaki. It is located in M101 (the pinwheel galaxy) at a distance of 6.4 Mpc at RA, Dec. = 210.9$^\circ$, 54.3$^\circ$ and is the closest and brightest supernova in nearly a decade \cite{discovery}. Several observatories, including the IceCube Neutrino Observatory, performed rapid follow-ups after the initial detection \cite{2023ATel16043....1T}.  

We describe a search for neutrinos coincident with SN 2023ixf using IceCube's Northern Tracks data sample and kernel density estimator method (KDEs) for improved energy and angular reconstruction \cite{Bellenghi:2023RX}. Section 2 describes the IceCube Neutrino Observatory and the data selection used in this analysis. Section 3 discusses models for neutrino production from SN 2023ixf, and section 4 discusses our search for neutrinos from the supernova. Finally, in section 5 we provide a discussion of this search as well as a comparison of previous IceCube supernova analyses.  

\begin{figure}[h]
    \centering
    \includegraphics[width=0.6\textwidth]{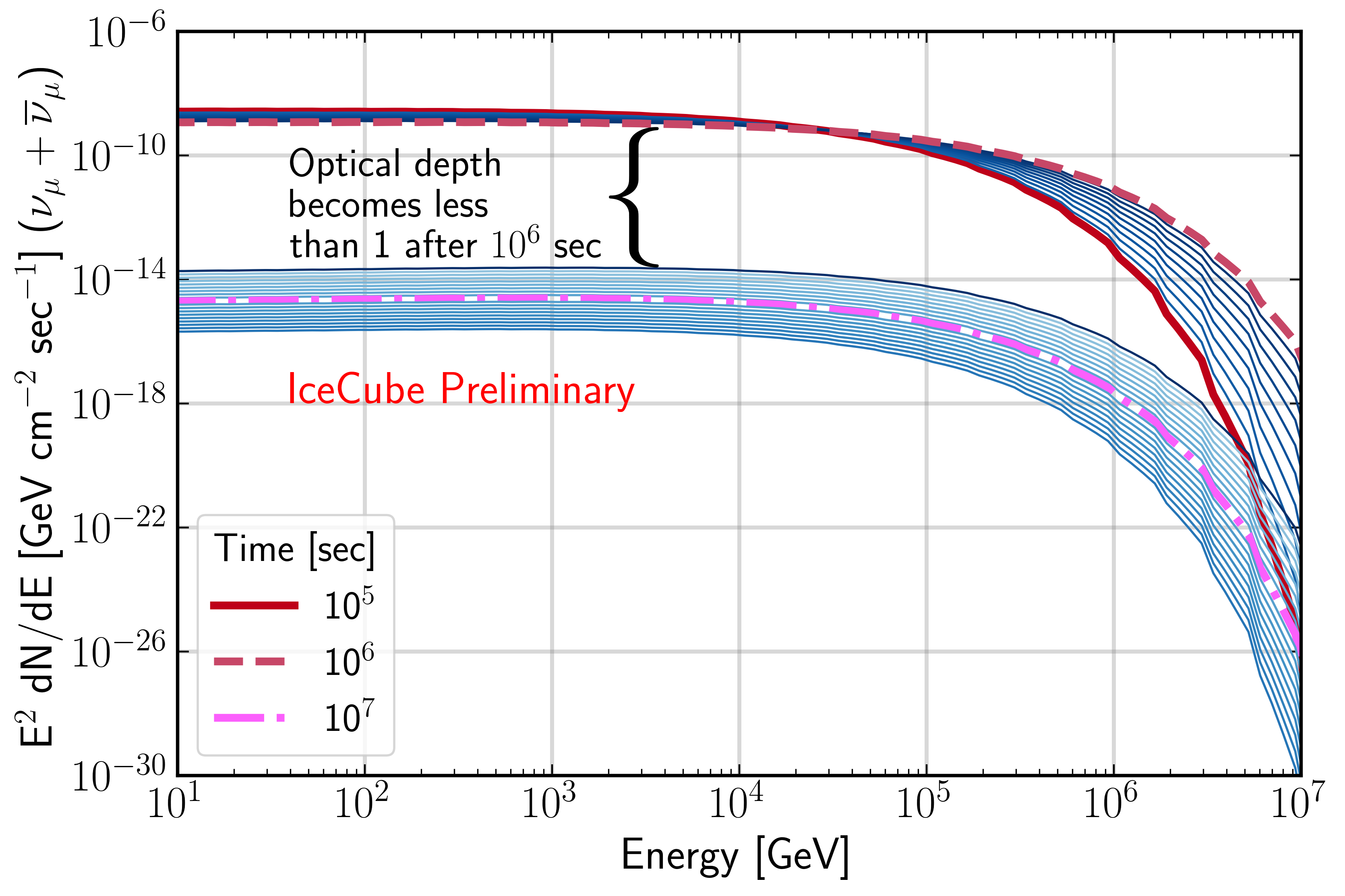}
    \caption{Time evolution of predicted neutrino flux from SN 2023ixf according to \cite{kheirandish_detecting_2023}. After approximately $10^6$ seconds, the optical depth becomes less than 1, significantly reducing the flux. The blue lines correspond to different instances in time with spacing of 10 lines per decade. At high energies, the availability of simulation decreases per energy decade, resulting in the "wiggle" features \cite{PhysRevD.97.081301}.}
    \label{fig:timeEvolution}
\end{figure}

\section{Models for Neutrino Production}\label{sec3}
During the gravitational collapse of the core of a massive star ($M_* > 8M_\odot$) during a type-II supernova, most of the gravitational binding energy is converted into thermal neutrinos with energies on the order of tens of MeV. In addition to these neutrinos, supernova progenitors surrounded by dense circumstellar material (CSM) have been predicted to produce high-energy neutrinos. Specifically, supernova ejecta interacts with the CSM and acts as a proton accelerator through diffusive shock acceleration. The accelerated protons interact with other protons present in the CSM and produce high-energy neutrinos via the p-p interaction. These neutrinos, which can have energies of up to TeV scale, would be detectable by IceCube \cite{kheirandish_detecting_2023, zirakashvili_type_2016, sarmah_high_2022, pitik_optically_2023}.

In order to motivate our analysis, we examine the model presented in \cite{kheirandish_detecting_2023} for neutrino production in SN 2023ixf through proton acceleration in CSM. In this model, neutrino flux is a function of the time after the supernova explosion. After a time period of approximately $10^6$ seconds, neutrino production slows down as the optical depth becomes less than 1, which is shown in Figure \ref{fig:timeEvolution}.

\begin{figure}
    \centering
    \includegraphics[width=0.6\textwidth]{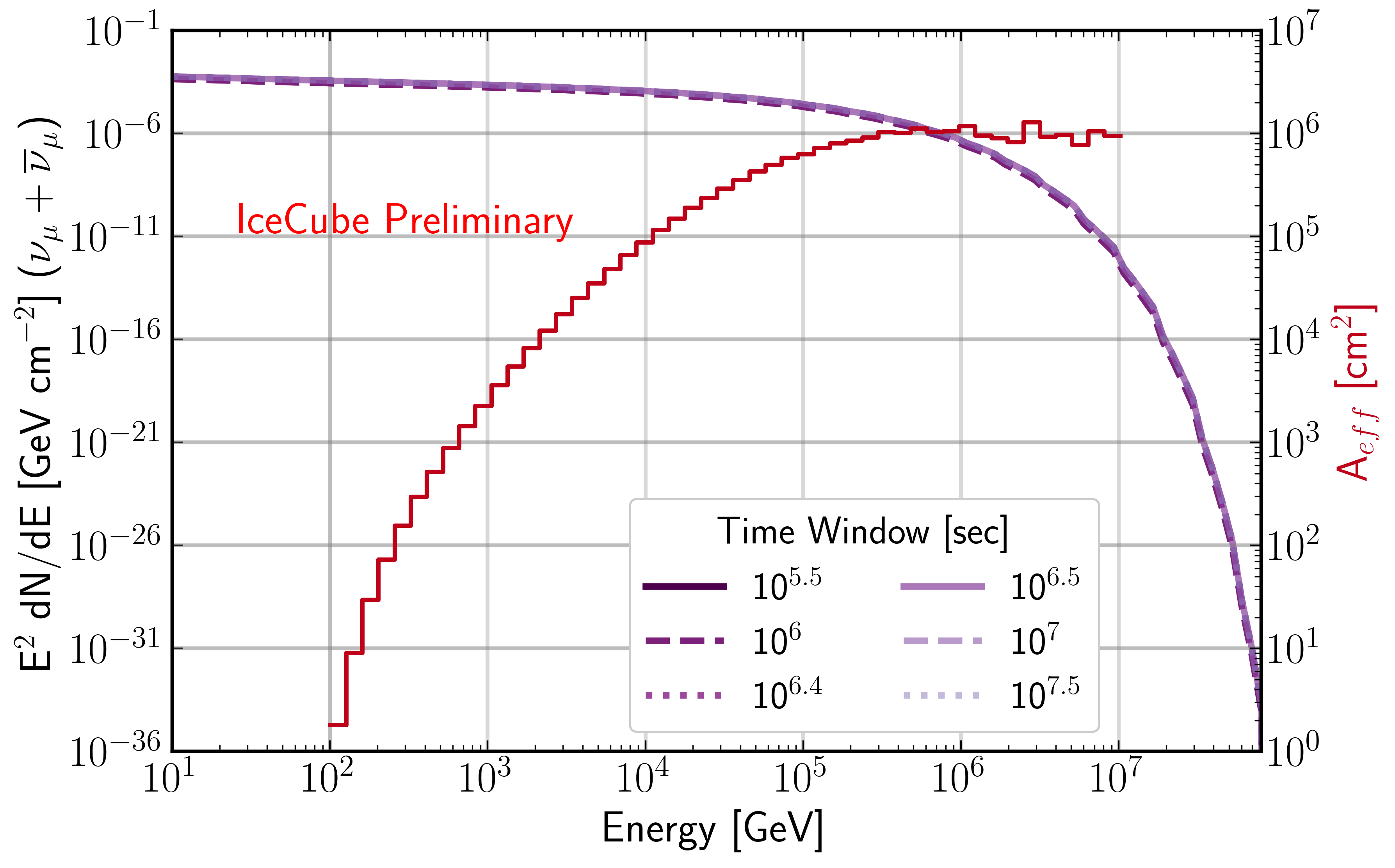}
    \caption{Effective area of IceCube (red) as a function of neutrino energy (for $\nu_\mu + \overline{\nu}_\mu$) with modeled time-integrated flux from \cite{kheirandish_detecting_2023}. In this figure, all of the purple lines lie ontop of one another.}
    \label{fig:Aeff}
\end{figure}

\begin{figure}[h]
    \centering
    \includegraphics[width=0.55\textwidth]{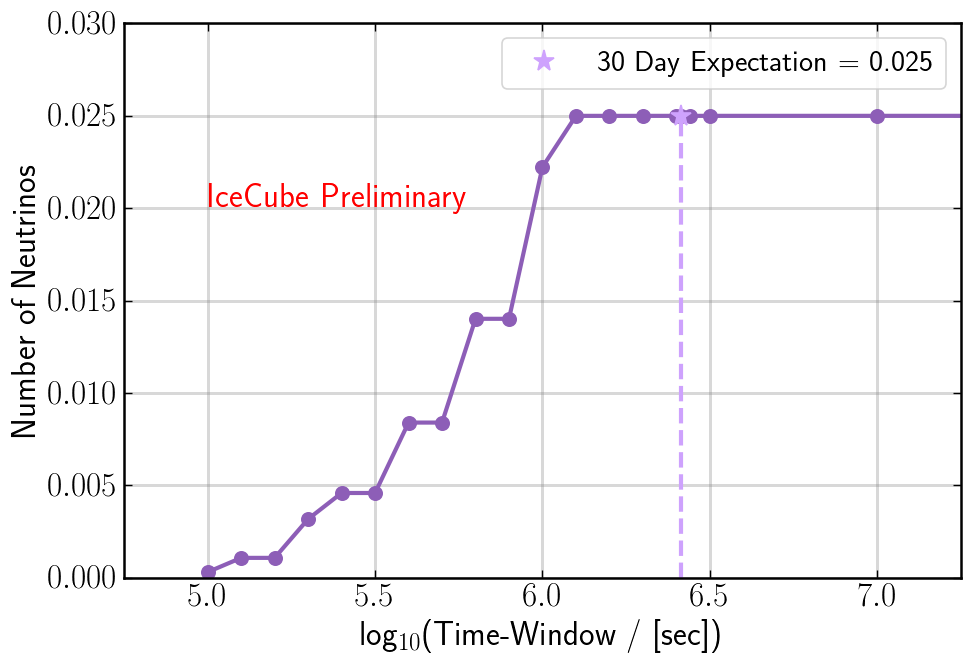}
    \caption{Model expectation of neutrinos detected by IceCube from SN 2023ixf as a function of time after supernova explosion. There is no significant increase in neutrinos after approximately $10^{6.1}$ sec. Our selected time-window of 30 days post-explosion has been marked with a star -- which has an expectation of 0.025 detected neutrinos. The step behavior is a feature of the time binning in the model presented in \cite{kheirandish_detecting_2023}.}
    \label{fig:Modelexpectation}
\end{figure}

We can use this model  to calculate the expected number of neutrinos detectable by IceCube by combining IceCube's effective area with the neutrino flux to produce expected neutrinos as a function of time after supernova explosion as seen in figure \ref{fig:Aeff}. The total time that we observe the supernova for is called a time-window. For a plot of the expected number of neutrinos as a function of time post-explosion, see Figure \ref{fig:Modelexpectation}.

\section{Neutrinos with IceCube}\label{sec2}
The IceCube Neutrino Observatory is a cubic kilometer neutrino detector located at the geographic South Pole consisting of 86 strings with a total of 5160 digital optical modules (DOMs). These DOMs are situated between 1.5 km and 2.5 km below the glacial surface in a hexagonal grid and are optimized for the detection of high-energy (TeV - PeV) neutrinos \cite{Aartsen:2016nxy}. The DOMs consist of 10-inch photomultiplier tubes encased in a pressurized glass housing for detecting Cherenkov emission from secondary charged particles from the interaction of neutrinos in the Antarctic ice. Neutrino detections result in two topologies of events: tracks and cascades. Charged current $\nu_\mu$ interactions product a track-like signature from the secondary muon that can travel up to several kilometers through the ice depending on the energy. All other neutrino interactions produce a near-isotropic light distribution in the detector called a cascade. Track-like signatures have an angular resolution of $\sim$1$^\circ$ and cascades have a worse resolution of $\sim$10$^\circ$.

For our analysis, we use the Northern Tracks data sample which contains track events in a declination range of $-5^\circ< \delta < 90^\circ$ and boasts a livetime of 13.5 years covering the period 06/01/2010 - 11/28/2023 (which includes the supernova discovery time). We select this event sample because our target declination ($\delta = 54.3^\circ$) is located in the northern hemisphere and Northern Tracks is the most sensitive dataset for the search of point sources in the northern hemisphere. Finally, kernel distribution estimators (KDEs) are implemented for the probability density functions (PDFs), which improves sensitivity by replacing the usual Gaussian approximation for the spatial PDF by a non-parametric inference of the same PDF \cite{Bellenghi:2023RX}.

\section{Follow-up}
\begin{figure}[h!]
        \centering
        \includegraphics[width=0.6\textwidth]{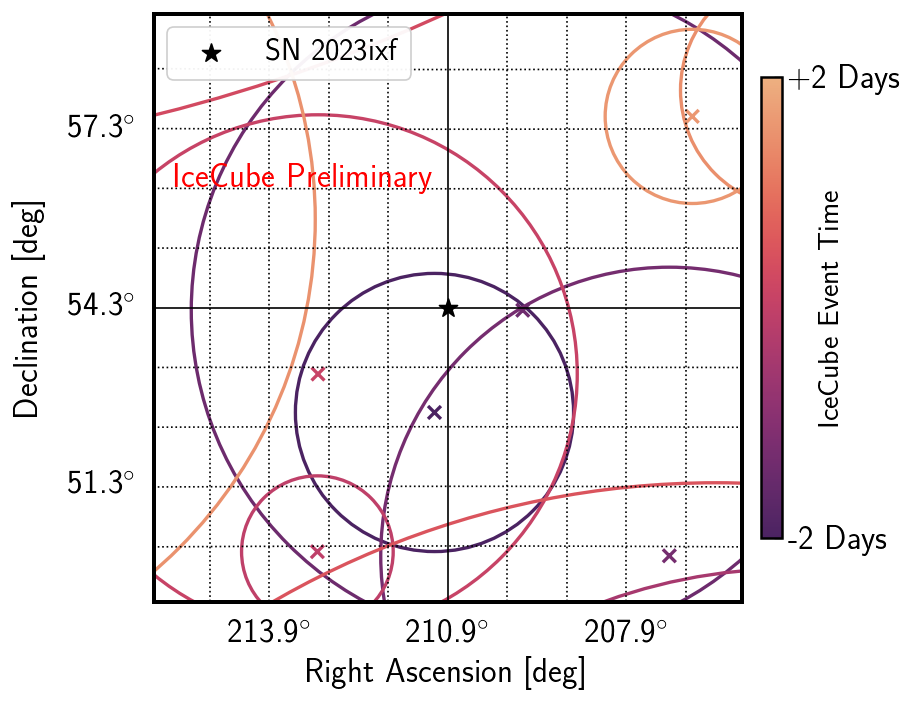}
        \caption{Fast response analysis zoomed-in skymap at the location of SN 2023ixf for a 4 day time window centered at ZTF discovery time. This depicts on-time events within $\pm$ 10 degrees of the target location in both RA and dec. IceCube obtained a p-value of 0.18, consistent with background expectation and an upper limit of E$^2$ dN/dE = 7.3 x 10$^{-2}$ GeV cm$^{-2}$ at 90\%CL for energy range between 600 GeV and 250 TeV. Here, the size of each circle corresponds to the angular uncertainty (90\%) for each neutrino.}
        \label{fig:skymap}
    \end{figure}

IceCube's realtime response of SN 2023ixf consisted of a 4-day time-window centered around the discovery time of the transient using IceCube's Gamma-ray Follow-Up (GFU) event sample for muon neutrinos above 100 GeV in energy. GFU is IceCube's online event sample and is used in realtime and fast response analyses (FRAs) \cite{abbasi_follow-up_2021}. IceCube obtained results consistent with background expectation and set time-integrated flux upper limits (E$^2$ dN/dE) of 7.3 x 10$^{-2}$ GeV cm$^{-2}$ assuming an E$^{-2}$ spectrum \cite{2023ATel16043....1T}. The zoomed-in skymap for the fast response analysis can be seen in Figure \ref{fig:skymap}. 
   
\begin{figure}[h]
    \centering
    \includegraphics[width=0.6\textwidth]{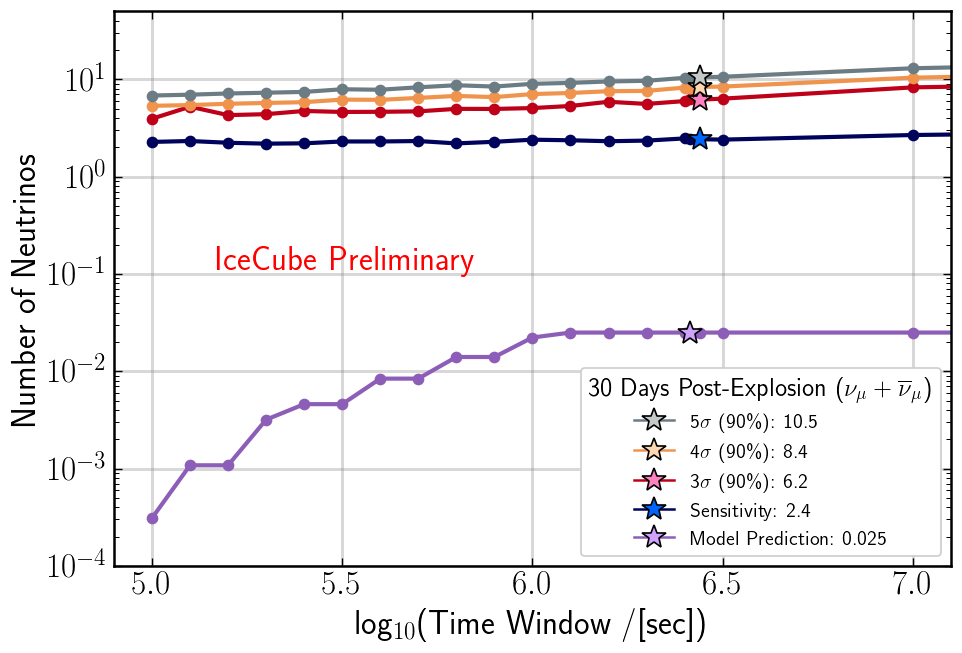}
    \caption{Sensitivity, $3\sigma, 4\sigma, 5\sigma$, and model expectation (taken from \cite{kheirandish_detecting_2023}) as a function of time-window. Target time window of 30 days post-explosion has been marked with a star for each line. This plot is for $\nu _{\mu }+{\overline {\nu }}_{\mu }$ only.}
    \label{fig:expectation}
\end{figure}

Using the model \cite{kheirandish_detecting_2023} discussed in section \ref{sec3} and the sensitivities of IceCube, we estimate the expected number of neutrinos for various time-windows in order to optimize our analysis. We see that after 30 days, there is not a significant increase in the number of expected neutrinos from SN 2023ixf (see Figure \ref{fig:expectation}). As a result, we are motivated to optimize our analysis for 30-days post discovery time. Using forced photometry data taken by the Zwicky Transient Facility (ZTF) to capture the last observation of SN 2023ixf pre-discovery, we include 2 additional days in our time window, prior to SN 2023ixf's discovery, in order to capture the explosion time. This gives us a total time-window of 32-days, with 2 days prior to discovery and 30 days post-discovery. 

We then construct a test statistic distribution using the following definition: 

\begin{equation}
\mathrm{TS}=-2 n_s+2\sum_{i=1}^N  \log \left[1+\frac{n_s S\left(\mathbf{x}_i, \mathbf{x}_s\right)}{n_b B\left(\mathbf{x}_i\right)}\right],
\end{equation}
where $n_s$ is the fitted number of signal events, $\mathbf{x}_i$ are the reconstructed properties of IceCube neutrino candidates, $\mathbf{x}_s$ is the location on the sky of the tested source position, $S$ is the signal probability density function (PDF), $B$ is the background PDF, $n_b$ is the expected number of background events, and $N$ is the total number of events in the event selection. This equation includes a Poisson term, which compares the total number of observed neutrinos  in a time-window with the number expected from background. We then maximize this test statistic only using events that occur within our selected time window of 32 days the model prediction and estimation of the explosion time from the optical light curve. For our analysis, we use a signal PDF which includes an energy PDF. For the energy PDF we assume that neutrino emission follows a power-law spectrum with an index of 2.0 to match the model presented in \cite{kheirandish_detecting_2023}. 

\begin{figure}
    \centering
    \includegraphics[width=0.6\textwidth]{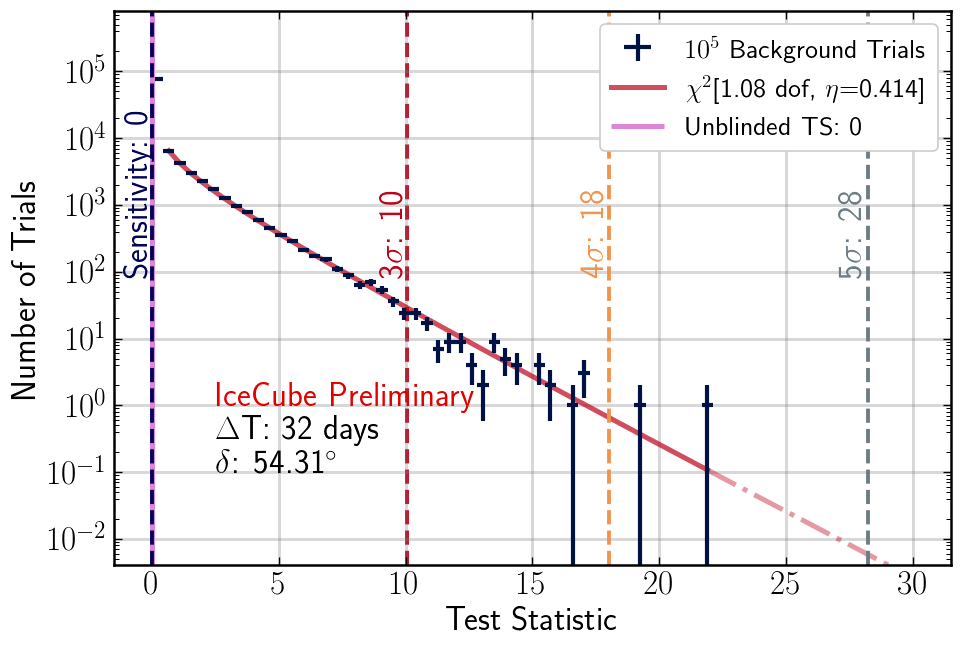}
    \caption{Test statistic distribution for SN 2023ixf at a time-window of 32 days, of which, 2 days are prior to discovery of the supernova. Unblinded TS of 0 is shown at the left side of the plot in a solid pink line, additionally we show the sensitivity of IceCube at the declination of SN 2023ixf and the 3$\sigma$, 4$\sigma$, and 5$sigma$ potentials. Here $\eta$ is the fraction of trials outside the 0 bin. At a TS of approximately 21, we no longer have background trials to fit $\chi^2$ to. We represent this extrapolation of the $\chi^2$ fit as a dash-dot line. For short time windows, if $\eta <5$, more than half of the background trials are in the TS = 0 bin, resulting in a median value of 0. Since we define our sensitivity as the number of neutrinos that result in 90\% of pseudoexperiments exceeding the median, our sensitivity TS is also 0.}
    \label{fig:TSDist}
\end{figure}

We calculate a test statistic distribution assuming a background only hypothesis. This is done by performing $10^5$ pseudoexperiments of data at our target declination that has been randomly shuffled in RA \cite{1993NIMPA.328..570A}. For each pseudoexperiment, we calculate the test statistic and combine all of them into a single test statistic distribution comprised of background-only. When we unblind our analysis, we can calculate the test statistic of our signal to obtain a signal test statistic. Comparing this to our background test statistic gives us the significance of our signal. A figure showing our background test statistic distribution in addition to our unblinded signal test statistic, sensitivity, and various levels of significance are given in Figure \ref{fig:TSDist}.

\section{Results and Discussion}
\begin{figure}[h]
        \centering
        \includegraphics[width=0.6\textwidth]{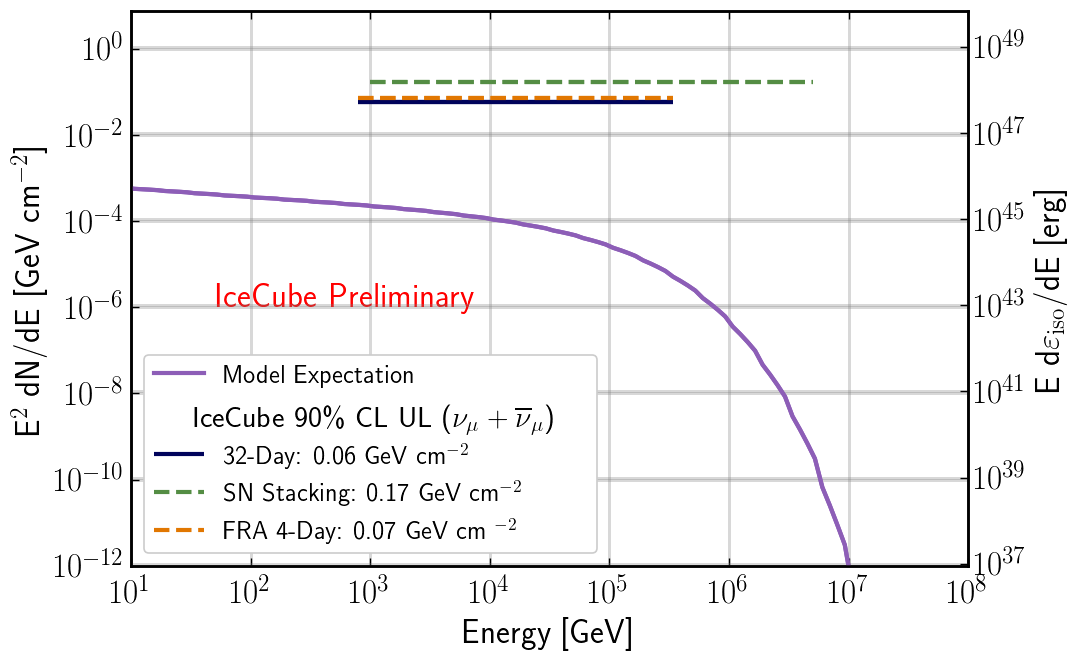}
        \caption{Upper limits from IceCube for SN 2023ixf. Included in this plot is the model expectation from \cite{kheirandish_detecting_2023} (purple), this analysis (dark blue), a previous IceCube Supernova Stacking analysis (green) \cite{abbasi_constraining_2023}, and the original IceCube SN 2023ixf Fast Response Analysis (orange) \cite{2023ATel16043....1T}. Here, the SN stacking line has been scaled for the distance to SN 2023ixf (distance = 6.4 Mpc) in order to make it more comparable to SN 2023ixf. The right axis shows the differential isotropic equivalent energy obtained with the expression: $E \frac{d\epsilon_{iso}}{dE} = \phi \frac{4\pi d_L^2}{1+z} $where $\phi = E^2 dN/dE$ and the appropriate unit conversions have been applied.
}
        \label{fig:UpperLimits}
    \end{figure}

    We present the results of a 32 day follow-up of SN 2023ixf. We obtain values consistent with background expectation and set strong upper-limits on the neutrino flux of a type-II core-collapse supernovae at our target declination of 0.06 GeV cm$^{-2}$. This limit represents the strongest limit on the high-energy neutrino flux from core-collapse SNe obtained using IceCube data. We present and summarize limits from 3 IceCube analyses, including this one, and set the strongest limits thus far for supernovae. Our results are summarized in Table \ref{tab:results}. These analyses include the initial realtime response to SN 2023ixf which encompasses a 4-day time-window centered at the supernova's discovery time, this analysis which uses a 32 day time-window with 2 days prior to discovery time and 30 days after, and finally, a stacking search for high-energy neutrino emission from core-collapse supernovae using IceCube data \cite{2023ATel16043....1T, abbasi_constraining_2023}. Additionally, we plot our upper limits and model expectation in Figure \ref{fig:UpperLimits}.
    \begin{table}[h]
        \centering
        \begin{tabular}{|l|c|} \hline
            Analysis & Upper-Limits (GeV cm $^{-2}$) \\ \hline \hline
            SN 2023ixf Follow-up (this work)& 0.06 \\ 
            Supernova Stacking Paper \cite{abbasi_constraining_2023} & 0.17 \\ 
            SN 2023ixf 4-Day Follow-up \cite{2023ATel16043....1T} & 0.07\\ \hline
        \end{tabular}
        \caption{Results from IceCube Supernovae Analysis}
        \label{tab:results}
    \end{table}

We expect long-term improvements to neutrino detection using the IceCube Neutrino Observatory with the implementation of the IceCube Upgrade, currently being installed at the South Pole. The IceCube Upgrade will extend IceCube's neutrino sensitivity to even lower energies (GeV and below) by increasing the density of the infill array's (IceCube DeepCore) strings \cite{Ishihara:2019uL}. On a more long-term timescale, IceCube Gen-2 will provide increased instrument volume and the addition of Gen-2 Radio which, when combined, will extend improved sensitivity to the TeV-PeV energy range for high-energy neutrino emission from nearby supernovae \cite{IceCube-Gen2:2020qha}. 

\bibliographystyle{ICRC}
\bibliography{references}

\clearpage

\section*{Full Author List: IceCube Collaboration}

\scriptsize
\noindent
R. Abbasi$^{16}$,
M. Ackermann$^{63}$,
J. Adams$^{17}$,
S. K. Agarwalla$^{39,\: {\rm a}}$,
J. A. Aguilar$^{10}$,
M. Ahlers$^{21}$,
J.M. Alameddine$^{22}$,
S. Ali$^{35}$,
N. M. Amin$^{43}$,
K. Andeen$^{41}$
% ...
\\
\\
% ...
$^{10}$ Universit{\'e} Libre de Bruxelles, Science Faculty CP230, B-1050 Brussels, Belgium \\
% ...
$^{16}$ Department of Physics, Loyola University Chicago, Chicago, IL 60660, USA \\
$^{17}$ Dept. of Physics and Astronomy, University of Canterbury, Private Bag 4800, Christchurch, New Zealand \\
% ...
$^{21}$ Niels Bohr Institute, University of Copenhagen, DK-2100 Copenhagen, Denmark \\
$^{22}$ Dept. of Physics, TU Dortmund University, D-44221 Dortmund, Germany \\
% ...
$^{35}$ Dept. of Physics and Astronomy, University of Kansas, Lawrence, KS 66045, USA \\
% ...
$^{39}$ Dept. of Physics and Wisconsin IceCube Particle Astrophysics Center, University of Wisconsin{\textemdash}Madison, Madison, WI 53706, USA \\
% ...
$^{41}$ Department of Physics, Marquette University, Milwaukee, WI 53201, USA \\
% ...
$^{43}$ Bartol Research Institute and Dept. of Physics and Astronomy, University of Delaware, Newark, DE 19716, USA \\
% ...
$^{63}$ Deutsches Elektronen-Synchrotron DESY, Platanenallee 6, D-15738 Zeuthen, Germany \\
$^{\rm a}$ also at Institute of Physics, Sachivalaya Marg, Sainik School Post, Bhubaneswar 751005, India \\
% ...

\subsection*{Acknowledgments}

\noindent
The authors gratefully acknowledge the support from the following agencies and institutions:
USA {\textendash} U.S. National Science Foundation-Office of Polar Programs,
U.S. National Science Foundation-Physics Division,
U.S. National Science Foundation-EPSCoR,
U.S. National Science Foundation-Office of Advanced Cyberinfrastructure,
Wisconsin Alumni Research Foundation,
Center for High Throughput Computing (CHTC) at the University of Wisconsin{\textendash}Madison,
Open Science Grid (OSG),
Partnership to Advance Throughput Computing (PATh),
Advanced Cyberinfrastructure Coordination Ecosystem: Services {\&} Support (ACCESS),
Frontera and Ranch computing project at the Texas Advanced Computing Center,
U.S. Department of Energy-National Energy Research Scientific Computing Center,
Particle astrophysics research computing center at the University of Maryland,
Institute for Cyber-Enabled Research at Michigan State University,
Astroparticle physics computational facility at Marquette University,
NVIDIA Corporation,
and Google Cloud Platform;
Belgium {\textendash} Funds for Scientific Research (FRS-FNRS and FWO),
FWO Odysseus and Big Science programmes,
and Belgian Federal Science Policy Office (Belspo);
Germany {\textendash} Bundesministerium f{\"u}r Forschung, Technologie und Raumfahrt (BMFTR),
Deutsche Forschungsgemeinschaft (DFG),
Helmholtz Alliance for Astroparticle Physics (HAP),
Initiative and Networking Fund of the Helmholtz Association,
Deutsches Elektronen Synchrotron (DESY),
and High Performance Computing cluster of the RWTH Aachen;
Sweden {\textendash} Swedish Research Council,
Swedish Polar Research Secretariat,
Swedish National Infrastructure for Computing (SNIC),
and Knut and Alice Wallenberg Foundation;
European Union {\textendash} EGI Advanced Computing for research;
Australia {\textendash} Australian Research Council;
Canada {\textendash} Natural Sciences and Engineering Research Council of Canada,
Calcul Qu{\'e}bec, Compute Ontario, Canada Foundation for Innovation, WestGrid, and Digital Research Alliance of Canada;
Denmark {\textendash} Villum Fonden, Carlsberg Foundation, and European Commission;
New Zealand {\textendash} Marsden Fund;
Japan {\textendash} Japan Society for Promotion of Science (JSPS)
and Institute for Global Prominent Research (IGPR) of Chiba University;
Korea {\textendash} National Research Foundation of Korea (NRF);
Switzerland {\textendash} Swiss National Science Foundation (SNSF).

\end{document}